\documentclass{ieeeaccess}

\usepackage{cite}
\usepackage{enumitem}
\usepackage{amsmath,amssymb,amsfonts}
\usepackage{algorithmic}
\usepackage{graphicx}
\usepackage{textcomp}
\usepackage{color}
\usepackage{listings}

\def\BibTeX{{\rm B\kern-.05em{\sc i\kern-.025em b}\kern-.08em
    T\kern-.1667em\lower.7ex\hbox{E}\kern-.125emX}}
\begin{document}
\history{Date of publication xxxx 00, 0000, date of current version xxxx 00, 0000.}
\doi{10.1109/ACCESS.2017.DOI}

\title{Critical Infrastructure Protection: Generative AI, Challenges, and Opportunities}
\author{\uppercase{Yagmur Yigit} \authorrefmark{1},
\IEEEmembership{Student Member, IEEE},
\uppercase{Mohamed Amine Ferrag} \authorrefmark{2}, \IEEEmembership{Senior Member, IEEE}, 
\uppercase{Iqbal H. Sarker}\authorrefmark{3},
\IEEEmembership{Member, IEEE}, 
\uppercase{Leandros A. Maglaras}\authorrefmark{1}, \IEEEmembership{Senior Member, IEEE}, 
\uppercase{Christos Chrysoulas}\authorrefmark{1}, \IEEEmembership{Member, IEEE}, 
\uppercase{Naghmeh Moradpoor} \authorrefmark{1},
\IEEEmembership{Senior Member, IEEE},
\uppercase{Helge Janicke} \authorrefmark{3},
\IEEEmembership{Member, IEEE}
}
\address[1]{School of Computing, Edinburgh Napier University, 10 Colinton Road, Edinburgh. EH10 5DT, Edinburgh, UK}

\address[2]{Artificial Intelligence \& Digital Science Research Center, Technology Innovation Institute, UAE}
\address[3]{Centre for Securing Digital Futures, Edith Cowan University, Perth, WA-6027, Australia.}
\address[4]{Cyber Security Cooperative Research Centre (CSCRC), Edith Cowan University, Perth, WA 6027, Australia}
\tfootnote{This paragraph of the first footnote will contain support 
information, including sponsor and financial support acknowledgment. For 
example, ``This work was supported in part by the U.S. Department of 
Commerce under Grant BS123456.''}

\markboth
{Author \headeretal: Preparation of Papers for IEEE TRANSACTIONS and JOURNALS}
{Author \headeretal: Preparation of Papers for IEEE TRANSACTIONS and JOURNALS}

\corresp{Corresponding author: Leandros A. Maglaras (e-mail: l.maglaras@napier.ac.uk).}

\begin{abstract}
Critical National Infrastructure (CNI) encompasses a nation's essential assets that are fundamental to the operation of society and the economy, ensuring the provision of vital utilities such as energy, water, transportation, and communication. Nevertheless, growing cybersecurity threats targeting these infrastructures can potentially interfere with operations and seriously risk national security and public safety. In this paper, we examine the intricate issues raised by cybersecurity risks to vital infrastructure, highlighting these systems' vulnerability to different types of cyberattacks. 
We analyse the significance of trust, privacy, and resilience for Critical Infrastructure Protection (CIP), examining the diverse standards and regulations to manage these domains. We also scrutinise the co-analysis of safety and security, offering innovative approaches for their integration and emphasising the interdependence between these fields. Furthermore, we introduce a comprehensive method for CIP leveraging Generative AI and Large Language Models (LLMs), giving a tailored lifecycle and discussing specific applications across different critical infrastructure sectors. Lastly, we discuss potential future directions that promise to enhance the security and resilience of critical infrastructures. This paper proposes innovative strategies for CIP from evolving attacks and enhances comprehension of cybersecurity concerns related to critical infrastructure.
\end{abstract}


\titlepgskip=-15pt

\maketitle

\section{Introduction}
\label{sec:introduction}
\PARstart{C}{ritical} National Infrastructure (CNI) consists of The resources of a nation that are essential to the smooth operation of its economy and society. This comprises all of the things that are necessary for day-to-day living, such as infrastructure, networks, people, equipment, systems, services, and information. 
CNI covers various sectors, including food and agriculture; transportation like railways and airports; water supply, including clean water and wastewater treatment; health components, which include hospitals, dental care, and ambulances; financial services like banking, taxes, and accounting; energy, which consists of oil and gas refineries and electric utilities; and information and telecommunications, like the internet and mobile networks \cite{sarker2024multi}.

It is crucial to protect CNI and ensure its reliability and cybersecurity because nations depend on its operation and consistency. Any disturbance to their operations could potentially devastate physical security, national security, economic wealth, safety, and health. This includes household and business destruction that may result in evacuations, business closures, financial losses, deaths, health hazards, and environmental impacts. 
The emergence of Industry 4.0 ~\cite{Intro1}, along with the increased connectivity of devices associated with CNI and the integration of traditional computer networks, has expanded the attack surface of these critical assets. The attacks on CNI have been an ongoing issue for decades, and they appear to be growing in number, frequency, and impact. 

For example, in December 2015 ~\cite{Intro2}, the world witnessed the first power outage caused by a cyber-attack. This attack, which began with a phishing attack, resulted from the BlackEnergy malware, a Trojan used for conducting distributed denial-of-service (DDoS) attacks, cyber espionage, and information destruction. It targeted utility companies in Ukraine, leaving hundreds without electricity for six hours. Moreover, cyberattacks have targeted water companies for twenty years ~\cite{Intro3}. For instance, in 2019 ~\cite{Intro4}, a water distribution company in Kansas (USA) experienced an attack by a former employee who gained remote control of the company’s information system and proceeded to temper with the drinking water treatment process. Furthermore, in 2021 ~\cite{Intro5}, there were attacks on water treatment infrastructure in Norway by ransomware named Ryuk. The hackers aimed to profit significantly by encrypting the company's files and demanding a ransom. Additionally, APT34 ~\cite{Intro6} serves as an example of Advanced Persistent Threats (APTs), as identified by FireEye researchers in 2017. APT34 specifically targeted government organizations and financial, energy, chemical, and telecommunications companies in the Middle East. Furthermore, APT28 ~\cite{Intro7}, a Russian group known as Fancy Bear, Pawn Storm, and Sednit, is another example of Advanced Persistent Threats (APTs) targeting CNI. It was identified by Trend Micro in 2014 and conducted attacks against military and government targets in Ukraine and Georgia, as well as NATO organizations and US defence contractors. 

Criminals and state-sponsored hackers are increasingly targeting CNI to disrupt society. They are probing for vulnerabilities, gathering intelligence, and exploiting individuals and systems for financial gain. Consequently, it is only a matter of time before a specific CNI becomes a direct target. The expectation that Industrial Control Systems (ICS) and CNI are completely secure, isolated, and immune to attacks is no longer valid. No industry or organization can consider itself completely safe. 

To provide cyber security and reliability for CNIs, there are various efforts that both governments and agencies can employ, including:
\begin{itemize}
    \item Implementing an All-Hazards approach to risk management, considering cyber and physical threats to critical infrastructure integrity.
    \item Integrating Incident Response (IR) strategies with Business Continuity Planning (BCP) to ensure seamless continuity of operations during and after security incidents.
    \item Adopting a Consequence Management approach to manage critical infrastructure failures' immediate and long-term impacts, including economic, societal, and environmental consequences.
    \item Regularly assessing the security status of CNIs and conducting penetration testing to identify vulnerabilities and weaknesses.
    \item Employing robust security mitigation measures such as intrusion detection systems, cryptography methods, firewalls, anti-virus software, and emerging security technologies like Blockchain, Artificial Intelligence (AI), and machine learning.
    \item Establishing and enforcing policies for maintaining and updating software and hardware periodically to mitigate vulnerabilities arising from outdated systems.
    \item Providing comprehensive cybersecurity training to staff to enhance awareness and preparedness against cyber threats.
    \item Enforcing robust cybersecurity policies and operating procedures, ensuring compliance with regulatory frameworks and industry standards.
    \item Encouraging international cooperation and coordination to address cross-border cyber threats effectively, including information sharing and joint response efforts.
    \item Collaborating with industry experts and sharing threat intelligence to stay ahead of emerging cyber threats and vulnerabilities.
\end{itemize}

Businesses can safeguard and preserve their critical infrastructure while guaranteeing that users will always have access to required services by putting these measures into practice. However, the tension between the requirement for information exchange and regulation and compliance must be acknowledged. Regulations are essential for establishing guidelines and guaranteeing accountability, but if they are burdensome, organizations are reluctant to report due to potential fines for law-breaking. Dealing with the cybersecurity challenges that CNIs encounter requires an open and cooperative culture, particularly in sectors where private businesses are common. Encouraging information exchange while maintaining regulatory control is necessary to achieve this. 

There is great potential for Critical Infrastructure Protection (CIP) when cutting-edge technologies like Large Language Models (LLMs) and generative AI are integrated. However, there are still a lot of hurdles to overcome in order to close the gap between theoretical advancements and practical applications. To do so, the difficulties and potential paths forward for protecting critical infrastructure systems need to be thoroughly examined. In this paper, we examine concerns about cybersecurity threats to critical infrastructure. We also investigate the several regulations and standards that manage privacy, trust, and resilience in relation to securing critical infrastructure. We provide an in-depth investigation of the co-analysis of safety and security, emphasizing the links between these fields while also offering innovative techniques for their integration. Furthermore, we describe a comprehensive method for utilizing generative AI and LLMs for CIP, provide an example lifecycle and discuss particular applications across multiple critical infrastructure industries. Finally, we suggest future paths that can enhance critical infrastructure security and resilience.  

The most popular reliability techniques are described in Section~\ref{sec:reliability}. An overview of the cybersecurity threats to critical infrastructure networks and their operations is provided in Section~\ref{sec:security}. Section~\ref{sec:privacy} delves into trust, privacy, and resilience requirements specific to CIP. Section~\ref{sec:securability} explores the interplay between safety and security and presents recent research advancements. In Section~\ref{sec:GenAI}, we delve into the practical applications of Generative AI and LLMs for enhancing critical infrastructure resilience and security. Section~\ref{sec:future} discusses future directions for CIP. Lastly, Section~\ref{sec:conclusion} summarizes the key findings and outlines potential advancements for future research and innovation in CIP.

\section{Reliability of Critical National Infrastructures}
\label{sec:reliability}
The reliability of CNIs is vital for critical networks and systems that operate normally since infrastructures support numerous businesses, including energy, transportation, telecommunications, and maritime ports. These industries offer critical services for a country's population's safety and well-being \cite{IEEE2022, EoT2023}. Reliability refers to the ability of these systems to do their responsibilities within set parameters and time restrictions \cite{ReliabilityAccess2021}. A system is considered reliable if it meets the requirements of the application and has a high probability of successful operation over a specific period of time. Building a reliable system involves understanding each component's overall dependability and how it interacts with other systems.

Installing security measures and offering reliable services to clients are the main responsibilities of the CNI. These components have the ability to recognize disturbances such as mistakes or cyberattacks that disrupt operations and respond accordingly \cite{maglaras2023BRIDGE}. Using statistical techniques, the CNI reliability examination analyzes the behaviour of the system and identifies any issues \cite{montecarlo}. This section covers the most widely used techniques for evaluating system reliability, such as Monte Carlo simulation methodologies, Weibull analysis, and Markov chains. In terms of system reliability, these are the best and most commonly used methods for defect analysis and system performance forecasts.

\subsection{Weibull Analysis}
The Weibull distribution methodology is the most efficient method for making inferences from failure data in components and systems. This method performs better even with small sample sizes than Poisson or binomial distribution methods \cite{weidist2023}. It is beneficial since collecting large failure samples would be very expensive and dangerous.

The Weibull analysis is a practical tool for modelling system behaviour with regard to reliability. Predictive reliability analysis has the advantage of being able to simulate several data sources. The approach most frequently employed is the two-parameter Weibull distribution, which is comprised of the scale parameter \(\alpha\) and the shape parameter \(\beta\). These components must be thoroughly understood in order to comprehend the failure process of a system. The failure rate function, average lifespan, reliability function, and likelihood of failure at any given time may be easier to compute thanks to this type of analysis \cite{weidist2023}.

Understanding the probability density function is necessary in order to calculate the predicted rate of failures over time. This function's expression is as follows:
\begin{equation}
    f(t; \alpha, \beta) = \frac{\beta}{\alpha} \left( \frac{t}{\alpha} \right)^{\beta - 1} e^{-\left( \frac{t}{\alpha} \right)^\beta}
\end{equation}

The likelihood that a failure will occur by a given time \(t\) may be found using the cumulative distribution function. As demonstrated by:
\begin{equation}
    F(t; \alpha, \beta) = 1 - e^{-\left( \frac{t}{\alpha} \right)^\beta}
\end{equation}

The failure rate function calculates the imminent failure risk, assuming the object has survived till time \(t\). It is described as
\begin{equation}
    \lambda(t; \alpha, \beta) = \frac{f(t; \alpha, \beta)}{1 - F(t; \alpha, \beta)} = \frac{\beta}{\alpha} \left( \frac{t}{\alpha} \right)^{\beta - 1}
\end{equation}

Failure intensity is mostly determined by the reliability function, which expresses the probability of surviving until at least time \(t\) \cite{Wei2011}. It gives a direct measure of survival, which is a supplement to the probability density function and is represented as:
\begin{equation}
    R(t; \alpha, \beta) = e^{-\left( \frac{t}{\alpha} \right)^\beta}
\end{equation}

The Weibull analysis relies heavily on the interaction between the shape (\(\beta\)) and scale (\(\alpha\)) parameters. Comprehending these parameters is imperative for precise reliability approximations, hence facilitating the refinement of maintenance tactics and product blueprints for amplified system effectiveness and dependability.
The shape parameter \(\beta\) determines the failure rate's trend over time, which indicates whether it rises, falls, or stays constant \cite{weiAccess2020}. On the other hand, the failure data's spread is impacted by the scale parameter \(\alpha\), which modifies the distribution's time axis \cite{weidist2022}. When these data are combined, it is possible to describe failure mechanisms precisely. It simplifies the process of building focused maintenance and replacement programs that boost system reliability.

\subsection{Markov Chains}
Markov Chains offer a mathematical framework for modelling and assessing the dependability of complexly interconnected and state-transitioning stochastic systems. Using Markov chains, the reliability analysis approach measures the likelihood of system states over time while taking into consideration all potential states of operation, failure, and repair procedures. This method makes it possible to evaluate the performance of the system and find reliability indices that are useful for operations and maintenance plans. 

\Figure[htbp](topskip=0pt, botskip=0pt, midskip=0pt){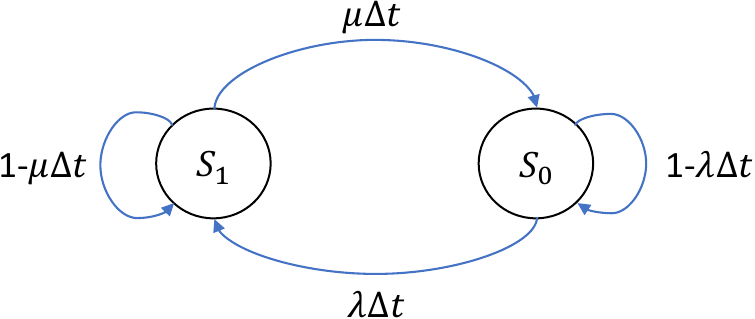}
{The representation of the transition of a two-state system.\label{fig:state-transition}}

A detailed examination of system reliability is achieved through a series of steps in the Markov Chain reliability analysis \cite{Markov2021}. These operations consist of:
\begin{itemize}
    \item Identifying every potential state of the system, including failure and normal states, to create a complete state space.
    \item Creating a transition probability matrix that shows the possibility of a state change at a given moment. This matrix is the foundation of Markov chain analysis.
    \item Indication of the rates of transition between the states, including the rates of failure and repair. For continuous-time Markov chains, it is usually assumed to follow exponential distributions.
    \item Determination of the long-term behaviour of the system by calculating the steady-state probability of each system state. This involves solving the balance equations derived from the transition probability matrix.
    \item Employing steady-state probabilities to determine key reliability metrics such as mean time to repair (MTTR), system availability, and mean time to failure (MTTF).
\end{itemize}

The state space \(S\) represents all possible states of the system, including operational and failure states.  
\begin{equation}
    S = \{s_0, s_1, s_2, \ldots, s_n\}
\end{equation}

The two-state system graph can be shown in Fig.~\ref{fig:state-transition}. In this diagram, \(s_0\) represent normal state, while \(s_1\) shows faulty state. The failure and repair rates are shown by  \(\lambda\) and \(\mu\), respectively. The time interval \(\Delta t\) depicts a brief time in Fig.~\ref{fig:state-transition} \cite{Markov2021}.

Markov chains predict future states from existing situations. This is highly beneficial for investigating the dependability of CNI. The definition of the transition probability matrix \(P\) is given by \(P_{ij}\), which denotes the likelihood of a single time step transfer from state \(s_i\) to state \(s_j\).

\begin{equation}
    P = \begin{bmatrix}
    P_{s_11} & P_{s_12} & \cdots & P_{s_1n} \\
    P_{s_21} & P_{s_22} & \cdots & P_{s_2n} \\
    \vdots & \vdots & \ddots & \vdots \\
    P_{s_n1} & P_{s_n2} & \cdots & P_{s_nn}
    \end{bmatrix}
\end{equation}

The steady-state probabilities \(\pi\) are calculated by decoding \(\pi P = \pi\) with the condition \(\sum_{s_i=1}^{n} \pi s_i = 1\) \cite{Markov2021}. These probabilities reflect the system's long-term behaviour, indicating the likelihood of being in each state after a large number of transitions.

Key reliability metrics such as MTTF and system availability can be derived from the steady-state probabilities \cite{maglaras2022mean}. For a system with states categorized into operational and failure states, MTTF can be estimated as follows \cite{MArkov2012}:
\begin{equation}
    MTTF = \sum_{s_i \in \text{{Operational States}}} \frac{1}{\lambda s_i}
\end{equation}
where \(\lambda s_i\) is the failure rate of state \(s_i\).

As the likelihood of the system being in an operational state at time \(t\) equals the total of the probabilities of all operational states, it is possible to calculate the reliability \(R(t)\) of the system \cite{Markov2023}.
\begin{equation}
    R(t) = \sum_{\text{operational states } i} \pi_i(t)
\end{equation}

System availability \(A(t)\), considering both operational and repair states, is calculated as \cite{Markov2021}:
\begin{equation}
A(t) = \frac{\mu}{\lambda + \mu}
\end{equation}

Assuming a simple two-state model. The transition probabilities are significantly influenced by the failure (\(\lambda\)) and repair (\(\mu\)) rates of the elements, directly affecting the system's reliability.

Several factors influence the reliability analysis utilizing Markov chains, including \(\lambda\), \(\mu\), \(P\), and \(S\) \cite{Markov2021}. The initial probability distribution among the states might affect how the system behaves in the short term and how well maintenance processes work.
The likelihood of state changes affects the system's overall performance and its capacity to sustain operational states over time. Increased failure rates make a failure scenario more likely, which lowers the system's overall reliability. Increased repair rates enhance reliability and availability by allowing the system to keep running and learn from its errors.

The reliability analysis is also impacted by the relationships between the system's numerous components because they alter the probability of a transition. Modeling these interactions consistently is necessary to achieve stable reliability values. A comprehensive and advanced method of comprehending a system's behavior throughout time is provided by the model. Transition probabilities, steady-state probabilities, and failure rates combined offer a thorough method for predicting system performance and indicating possible areas for development.
 
\subsection{Monte Carlo Simulation}
The Monte Carlo Simulation technique is becoming increasingly necessary in terms of reliability. It provides a dependable means of assessing the performance and reliability of complex systems. In many situations, the behaviour of complex systems can be predicted by statistical modelling and random sampling. This helps experts estimate the likelihood of system faults and identify important areas for improvement.
That makes it an invaluable technique for assessing risk \cite{Monte2008}. Utilizing probability distributions to depict the uncertainty in system performance and component dependability, this reliability analysis essentially models the system's behaviour under various scenarios. 
By simulating numerous situations in which each component may fail following its failure distribution and seeing the system's response to these failures, the reliability of a system is determined.

There are multiple important steps in the Monte Carlo Simulation process \cite{montecarlo, montecarlo2, monte2023}:
\begin{itemize}
    \item \textit{Definition of a System Parameter:} Set up initial system parameters ($P$), such as operating conditions ($\mu$), failure rates ($\lambda$), and repair rates.

    \item \textit{Setup of the Simulation Model:} Create a model, $S(t)$, that represents the system's operational states over time, with $S(t) = 1$ representing normal operation and $S(t) = 0$ representing failure.

    \item \textit{Requirements for Failure:} Define performance thresholds as the basis for failure criteria and designate the system as failed when its performance ($P$) falls below a given threshold ($P_{th}$).

    \item \textit{Stochastic Sampling:} For every parameter, perform a random sampling from the corresponding probability distributions; for example, sample $t_{fail}$ for failure rates from $Exp(\lambda)$.

    \item \textit{Iteration and Statistical Analysis:} Conduct multiple simulation iterations ($N$) to observe various outcomes, calculating the system reliability ($R$) and time to failure ($TTF$) as follows:
    \begin{equation}
        R = \frac{1}{N}\sum_{i=1}^{N}S_i(t),
    \end{equation}
    where $S_i(t)$ represents the system state in the $i$-th iteration at time $t$.

    \item \textit{Analysis of Results:} Estimate $MTTF$ and reliability over the specified period with:
    \begin{equation}
        MTTF = \frac{\sum_{i=1}^{N}TTF_i}{N},
    \end{equation}
    where $TTF_i$ is the time to failure in the $i$-th iteration.

\end{itemize}

The failure rate ($\lambda$), system architecture, performance thresholds ($P_{th}$), and operational environment in Monte Carlo Simulations for reliability analysis interplay significantly \cite{monte2023} \cite{Monte2008}. An increase in $\lambda$ typically reduces $R$ and $MTTF$, indicating lower system reliability. Adds complexity to $S(t)$ modelling, where redundancy can enhance $R$ but also introduces additional variables to the simulation. Modifying $P_{th}$ alters the failure criteria, affecting $R$ and $MTTF$ calculations. Variations in operational conditions can impact $\lambda$ and $\mu$, thereby influencing $R$ and $MTTF$.

The Monte Carlo Simulation technique is highly practical in investigating the impact of diverse risk factors on CNI and estimating the effectiveness of different mitigation tactics. It comprehensively understands potential vulnerabilities and resilience strategies, considering various inputs and outcomes. It provides insights into how different factors contribute to system reliability, guiding design and maintenance strategies decision-making.

\section{Cybersecurity issues}
\label{sec:security}
Cybersecurity issues significantly threaten critical infrastructure reliability, operation, and consistency. The interconnected nature of these systems and their reliance on digital technologies make them vulnerable to cyberattacks, which can have far-reaching effects on society.

One of the primary concerns regarding cybersecurity for critical infrastructures is the potential for malicious actors to infiltrate and disrupt essential systems. Cyber threats can range from simple phishing attempts to sophisticated malware injections and ransomware attacks \cite{maglaras2023BRIDGE}, which all compromise the integrity and functionality of critical infrastructure networks. For example, a successful attack on an energy grid could result in widespread power outages, affecting millions of individuals and businesses. The following are some common and highly prevalent cyber threats that CNIs should be aware of.

\begin{itemize}
    \item \textit{Malware:} Malicious software like viruses, worms, and Trojan horses can compromise the integrity, availability, and confidentiality of critical infrastructure systems. A malware program may be designed to steal sensitive data, disrupt operations, or allow attackers to take control of infrastructure assets remotely.
   
    \item \textit{Ransomware:} Critical infrastructure has increasingly been targeted by ransomware attacks. The attacks can disrupt operations and demand large ransom payments, resulting in financial losses and outages.

    \item \textit{Supply Chain Attacks:} A critical infrastructure often depends on third-party vendors for hardware, software, and services. The threat of supply chain attacks, where attackers compromise suppliers to gain access to target infrastructure, is becoming more common and difficult to detect.

    \item \textit{Phishing:} Phishing attacks target employees or system users in an attempt to obtain sensitive information, such as login credentials or financial information. By impersonating legitimate entities, such as utility providers and government agencies, phishing emails or messages can gain access to critical infrastructure networks.

    \item \textit{Denial-of-Service:} By overloading critical infrastructure systems with traffic, these attacks cause them to become slow or unresponsive. Multi-device DDoS attacks can disrupt essential services like communication networks and online utilities.

    \item \textit{SQL Injection:} Databases are targeted by SQL injection attacks that exploit vulnerabilities in web applications. Attackers can manipulate SQL queries to access, modify, or delete sensitive data stored in critical infrastructure systems.

    \item \textit{Zero-Day Exploits:} Zero-day exploits can take advantage of previously unknown vulnerabilities in software or hardware that have not yet been patched. These vulnerabilities are exploited by attackers to gain unauthorized access to critical infrastructure systems, steal data, or disrupt operations before security patches are available.
\end{itemize}

Moreover, a breach in one sector can cascade to others due to the interconnected nature of critical infrastructure. For instance, an attack on a transportation network could disrupt the supply chain, causing shortages of essential goods and services \cite{vanet2024}. The interconnectedness of our world amplifies cybersecurity issues and highlights the need for comprehensive protection measures.

Cybersecurity issues can also undermine the reliability and consistency of critical infrastructure operations. A security breach erodes trust in these systems, which are crucial for effective functioning. This may cause stakeholders to hesitate to rely on critical infrastructures, leading to disruptions in service delivery and economic instability. Furthermore, recovering from cyber attacks can be costly, further straining resources and disrupting operations. 

Overall, it is important to mitigate the risk of these attacks by prioritizing cybersecurity and implementing comprehensive protection measures. A failure to address these challenges effectively could have severe consequences for national security, economic stability, and societal well-being. 

\section{Trust, Privacy and Resilience}
\label{sec:privacy}
CNIs represent vital systems essential for the functioning of a nation or region, imposing the need to adhere to stringent privacy standards. The specific privacy, trust and resilience, in most cases, requirements applicable to CNIs vary based on location and characteristics and are subject to diverse regional standards, laws, and regulations. Examples include the General Data Protection Regulation (GDPR) in the European Union \cite{GDPR-EU}, the Health Insurance Portability and Accountability Act (HIPAA) in the United States \cite{HIPAA-US}, and the Personal Data Protection Act (PDPA) in Singapore \cite{PDPA-S}.

To safeguard CNIs, privacy requirements aim to ensure the integrity, confidentiality, and availability of associated information and systems. Key considerations include confidentiality and data protection, restricting system access to authorized personnel, compliance with privacy laws like GDPR and HIPAA, and implementing robust network security measures, such as blockchain and cryptography. Access to CNIs is secured through physical security measures like video surveillance and alarms. The resilience of CNI systems is assured through comprehensive recovery and data backup processes, employing techniques like data replication and cloud backup.

To fortify cybersecurity, CNIs deploy a spectrum of techniques encompassing network security, application security, data security, identity and access management, and risk management. These measures, including advanced encryption standards and innovative solutions proposed in the literature \cite{9798072, s22145105}, collectively contribute to a multi-layered defence against cyber threats. Periodic monitoring, audits, and a continuous commitment to privacy compliance are integral components to guarantee the efficacy of these security measures, fostering the overall resilience and reliability of CNIs.

\begin{figure*}[htbp]
    \centering
    \includegraphics[width=0.9\textwidth]{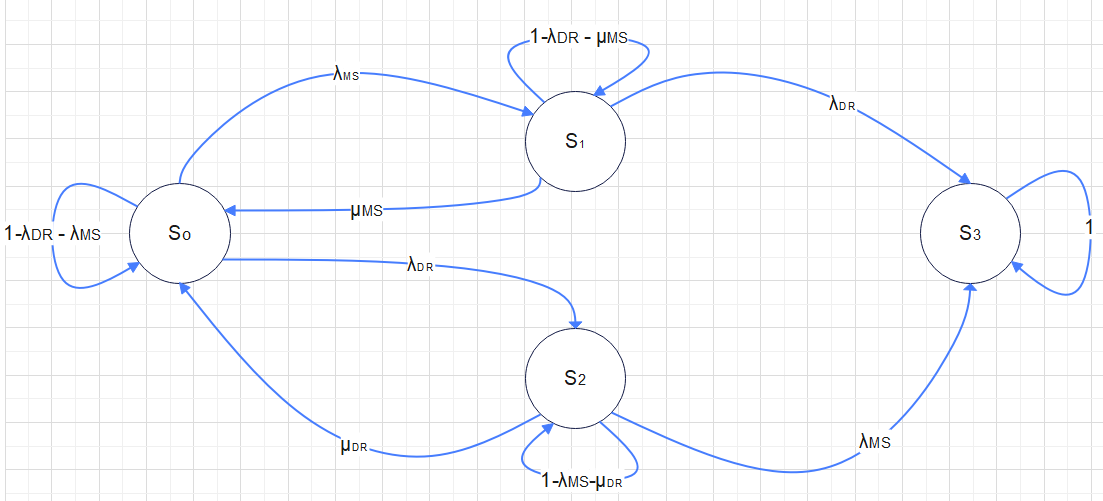}
    \caption{A Markov Chain of an MS/DR system}
    \label{fig:DR}
\end{figure*}

\section{Securability}
\label{sec:securability}
Scholars who conduct research in security or safety tend to address each field independently of the other. We strongly believe that these fields are interdependent, and based on some recent works, we present the current research in this area. The co-analysis of safety and security can be classified into two major categories: integrated strategy and unified strategy \cite{kavallieratos2020safesec}. The main difference between integrated strategy and unified strategy lies in the approach they take when they combine security and safety, the former focusing on integrating the results while the latter on the co-analysis of the system. As stated in \cite{fan2022safety}, safety and security co-analysis (SSCA) could benefit accident prevention in the transportation sector.  To differentiate between security attacks and safety problems, scholars classify the events that lead to threats or hazards. They also state that security risks come from deliberate actions while safety risks come from mistakes or errors \cite{lautieri2005safsec}, neglecting that mistakes of the users initiate many security attacks \cite{evans2019heart}.

The idea of including a probabilistic model of the behaviour of a part or the whole system in the form of suspected failures or faults could provide a better picture of the system in the analysis and a prediction of future states. For example, let us imagine that we are trying to analyze the behaviour of a system from a high-level perspective when the system also has a disaster recovery facility. For disaster recovery to work, the data and computer processing must be replicated at an off-site location that is unaffected by the incident. An organization needs to recover lost data from a backup location if the servers go down due to a natural disaster, equipment malfunction or cyber attack. To maintain operations, a business should also be able to move its computer processing to this remote location so that it can continue to provide its services to its customers.

The main system is represented as MS and the disaster recovery site as DR. If we work on an abstract level, we can represent the states of the system using a Markov chain, where:
\begin{itemize}
    \item State $S_o$ is when both the system and the DR site are operating normally.
    \item State $S_1$ is when the system is down due to a malfunction or attack.
    \item State $S_2$ is when the DR site is switched off.
    \item State $S_3$ is when both S and DR are switched off
\end{itemize}
where $\lambda_{MS}$ is the failure rate of the system, and $\lambda_{DR}$ is the failure rate of the disaster recovery site.

The transition from state $S_0$ or state $S_1$ to state $S_2$ occurs at rates $\mu_{MS}$ and $\mu_{DR}$ respectively, which represent the repair/recovery rate of the system/DR. 

For the organization to be able to offer the service around the clock, the $\lambda$ rates must be lower than the $\mu$ rates. Using the Markov model from Fig.~\ref{fig:DR}, we can calculate the MTTF (Mean Time to Failure) or MTTA (Mean Time to Attack) and the MTTR (Mean Time to Restore, Respond or Repair) depending on the model used. The correct values for these rates require a thorough analysis of the system components and their interdependencies, as well as an up-to-date assessment of the threats. This analysis is demanding and must be performed in several steps using a top-down approach. The main system can be divided into subsystems. A state transition diagram for each subsystem must be created, together with a general model that represents the dependencies between the subsystems in the general form r out of n (r out of n:G). In this model, at least the r subsystems or elements must be in a good state for the system to be operational. When incorporating cybersecurity into this reliability analysis, the calculation of the failure probability of each component must include failures as well as possible attacks.

\begin{figure*}[htbp]
    \centering
    \includegraphics[width=0.9\textwidth]{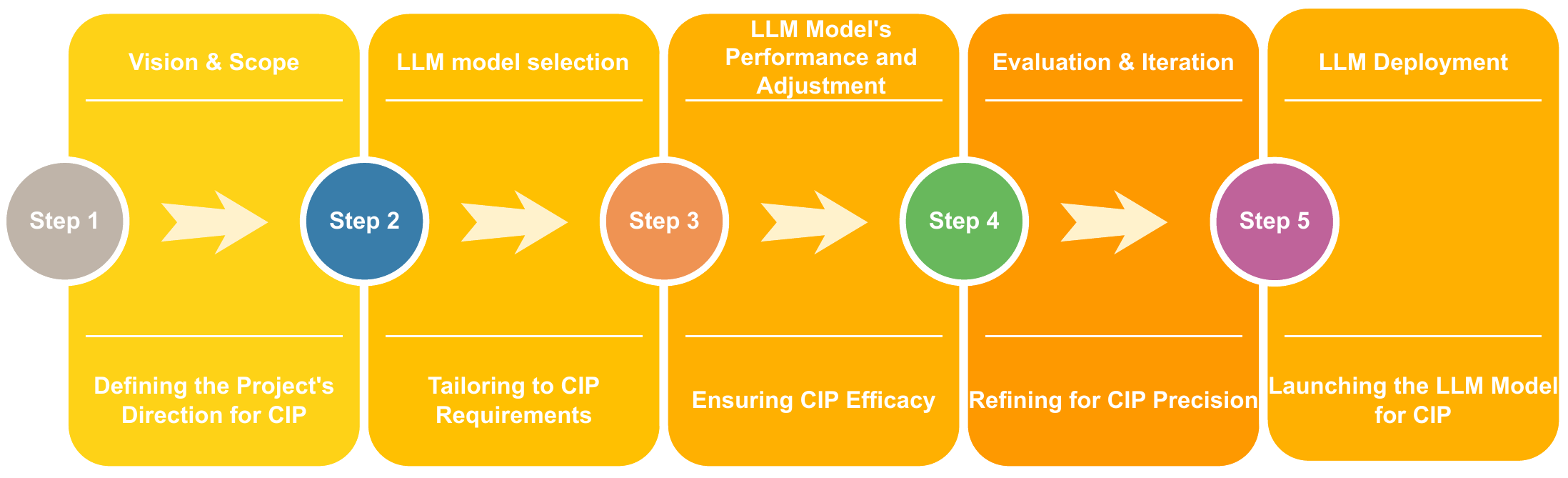}
    \caption{The steps of Generative AI and LLM lifecycle for critical infrastructure protection.}
    \label{fig:FigLLM1}
\end{figure*}

\section{Generative AI and Large Language Models for Critical Infrastructure Protection}
\label{sec:GenAI}
The deployment of Generative AI and LLMs in CIP is more than theoretical; it is a burgeoning reality with real-world applications that demonstrate the potential of these technologies \cite{alwahedi2024machine}. This section highlights specific examples of how LLMs have been utilized to enhance the resilience and security of critical infrastructure, from energy grids to water treatment facilities \cite{chowdhury2021cyber}.

\subsection{LLM lifecycle for Critical Infrastructure Protection}

For an application focused on CIP using LLMs, tailoring the lifecycle to emphasize CIP's unique challenges and requirements—such as security, resilience, and domain specificity—is critical \cite{zhu2024generative}. In this sub-section, we discuss the LLM lifecycle for CIP, which is based on the following five steps, as presented in Fig.~\ref{fig:FigLLM1}.

\subsubsection{Vision \& Scope: Defining the Project's Direction for CIP}
\begin{itemize}
    \item \textit{Objective Clarification:} We establish the model's role in protecting critical infrastructure. `Will it analyze threat intelligence, aid vulnerability assessments, or assist in emergency response?' Setting a clear, CIP-focused objective will guide the development process.
     \item \textit{Scope Determination:} We identify which critical infrastructure sectors the LLM will focus on, such as energy, water, and transportation. Different sectors may require different types of data and domain knowledge.
\end{itemize}

\subsubsection{Model Selection: Tailoring to CIP Requirements}
\begin{itemize}
    \item \textit{Security and Reliability:} We choose or develop a new model emphasizing security and data privacy, which are essential for CIP applications.
     \item \textit{Domain Adaptation:} We decide whether to adapt an existing LLM or train a new one with a dataset enriched with CIP-related content.
\end{itemize}

\subsubsection{Model's Performance and Adjustment: Ensuring CIP Efficacy}
\begin{itemize}
    \item \textit{Performance Assessment:} We evaluate the model's ability to identify, classify, and predict threats to critical infrastructure.
    \item \textit{Adjustment for CIP:} Focus adjustments on enhancing the model's capability to deal with the specific nuances of critical infrastructure threats. Hence, this could involve prompt engineering with CIP-specific prompts or further fine-tuning on targeted datasets.
\end{itemize}

\subsubsection{Evaluation \& Iteration: Refining for CIP Precision}
\begin{itemize}
    \item \textit{CIP-Specific Metrics:} We use evaluation metrics that reflect the model's performance in a CIP context—threat detection accuracy, response speed, and ability to work with domain-specific data.
\end{itemize}

\subsubsection{LLM Deployment: Launching the LLM Model for CIP}
Once deployed, we must establish mechanisms for ongoing monitoring of the model's effectiveness and updates to maintain its relevancy against evolving threats to critical infrastructure.

\subsection{Predictive Analysis and Threat Intelligence: The Case of Energy Grid Protection}
LLMs can be leveraged in the energy sector to detect potential cyber-attacks on power grids. For example, a company might utilize models like GPT-4 (Generative Pre-trained Transformer-4) to analyze and interpret extensive unstructured text data from online forums, threat reports, and system logs to predict and identify potential cyber threats, including phishing attacks or malware aimed at energy grid systems \cite{ferrag2024revolutionizing}. However, processing company data within an external data centre, such as OpenAI's, raises privacy concerns. Developing a sector-specific LLM model and deploying it within the company's data centre can enhance the protection of sensitive data.

\subsection{Automated Incident Response: Enhancing Pipeline Security}
BERT \cite{devlin2018bert}, known for its deep understanding of language context, is particularly useful for parsing and extracting specific information from incident reports, security logs, or communication between stakeholders involved in critical infrastructure. For example, a transportation authority could use BERT to quickly sift through incident reports following a security breach in a public transportation network, identifying common patterns or vulnerabilities that need immediate attention. Thus, it would speed up the response time and ensure the safety and reliability of transportation services.

\subsection{Enhancing Communication and Coordination: Water Treatment Facility Case Study}
T5 \cite{raffel2020exploring} can convert language-based tasks into a unified text-to-text format, making it exceptionally suitable for generating compliance and policy documentation vital for critical infrastructure sectors. A water treatment facility might leverage T5 to automate the creation of compliance reports based on new regulatory guidelines and operational data. This ensures accuracy and adherence to legal requirements and significantly reduces the administrative burden, allowing staff to focus on operational excellence and system integrity.

\subsection{Challenges and Considerations}
In the context of applying Generative AI and LLM models for critical infrastructure, integrating these models poses several open challenges. Each challenge requires careful consideration and innovative solutions to ensure the effective and secure application of LLMs. Below, we explore these challenges in more detail:

\subsubsection{Building an instruction Cyber Security dataset}
One of the primary challenges in leveraging LLMs for critical infrastructure is developing a comprehensive and relevant cybersecurity dataset. Critical infrastructure systems are highly complex and often proprietary, making it difficult to gather real-world data for training purposes. Additionally, the dataset must be diverse enough to cover various cyber threats and attack vectors unique to critical infrastructure sectors. Ensuring the dataset's quality, relevance, and privacy compliance also poses significant challenges, as it must be constantly updated to reflect evolving cyber threats. The structure of the Alpaca dataset can be adapted for building an instruction Cyber Security dataset  \cite{taori2023alpaca} as follows:

\begin{lstlisting}[language=, caption= Instruction Dataset Format]
Below is an instruction that describes a task, paired with an input that provides further context. Write a response that appropriately completes the request.
### Instruction:
{instruction}
### Input:
{input}
### Response:
\end{lstlisting}

\subsubsection{Pre-training models}
Pre-training LLMs for critical infrastructure applications also involves several challenges, including selecting appropriate pre-training tasks that align with cybersecurity contexts. The sheer volume of data required for effective pre-training and the computational resources needed are substantial. Moreover, the model must be trained to generalize across critical infrastructure sectors without compromising sector-specific requirements.

Developing pre-trained LLMs for CIP involves a meticulous process that starts with collecting and preparing high-quality, domain-specific datasets. In the context of critical infrastructure, this includes gathering extensive data from cybersecurity reports, threat intelligence feeds, and technical documents related to infrastructure systems. The preprocessing and cleaning of this dataset are vital steps, ensuring that irrelevant, duplicate, or sensitive information is removed, thereby refining the dataset to contain only the most relevant and high-quality data for training purposes \cite{driess2023palm,achiam2023gpt}.

Following the preparation of a meticulously curated dataset, the model architecture must be defined, considering critical infrastructure security's specific needs and challenges. This includes selecting the appropriate LLM architecture—such as variants of GPT or other autoregressive models—and adjusting hyperparameters to optimize performance for the unique context of critical infrastructure. The training involves teaching the LLM to predict the next word in a sequence and enabling it to understand complex cybersecurity concepts and the nuances of different threat vectors affecting critical infrastructure \cite{touvron2023llama,touvron2023llama2}.

\begin{figure*}[!ht]
    \centering
    \includegraphics[width=0.8\textwidth]{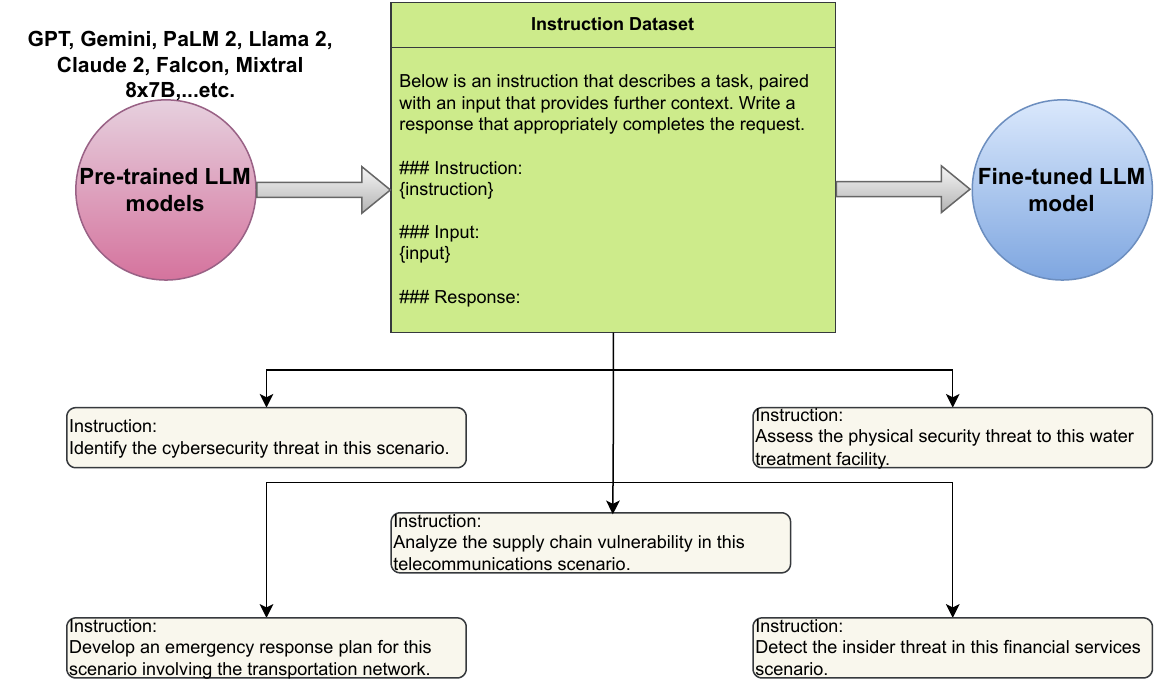}
    \caption{Instruction fine-tuning LLM models for Critical Infrastructure Protection.}
    \label{fig:FigLLM2}
\end{figure*}

\subsubsection{Supervised Fine-Tuning}
Supervised fine-tuning for LLMs in CIP involves updating pre-trained language models with specific, labelled datasets to perform targeted tasks more efficiently \cite{chang2023survey}, as presented in Fig.~\ref{fig:FigLLM2}. This process, distinct from unsupervised methods, enhances models' ability to interpret and react to nuanced requirements within the critical infrastructure domain. By employing labelled examples tailored to the unique challenges of infrastructure security, such as threat detection or system diagnostics, LLMs can offer more precise and relevant responses, improving overall security measures \cite{zhao2023survey}. Therefore, the limitation of supervised fine-tuning in applying LLMs for CIP lies in its reliance on high-quality, labelled datasets. This requirement can pose challenges in scenarios where such data are sensitive or expensive, potentially limiting the model's learning capability and adaptability to new or evolving threats within critical infrastructure sectors \cite{zhao2024explainability}.

Exploring fine-tuning methodologies beyond the conventional supervised approach can offer nuanced benefits and challenges. Transfer learning and task-specific fine-tuning stand out for their potential to adapt LLMs like GPT and BERT to specialized tasks, leveraging pre-existing vast datasets for efficiency and accuracy in targeted applications. However, these methods can also introduce risks such as catastrophic forgetting, where a model's performance on non-fine-tuned tasks deteriorates \cite{zhang2023instruction}.

Multi-task learning and sequential fine-tuning present solutions to broaden an LLM's capabilities across multiple tasks or gradually specialize its knowledge, mitigating the drawbacks of single-task focus \cite{min2023recent}. While demanding extensive datasets, these approaches enable the creation of versatile models capable of handling diverse tasks relevant to safeguarding critical infrastructure, thus offering a balanced strategy to exploit LLMs' strengths while addressing their limitations \cite{zhu2023large}.

\subsubsection{Reinforcement Learning from Human Feedback}
Reinforcement Learning from Human Feedback (RLHF) \cite{ouyang2022training} involves enhancing Generative AI and LLM by incorporating direct human feedback into the learning process. This method has significantly improved LLMs' relevance, accuracy, and ethical considerations, particularly in applications like chatbots. Integrating RLHF into LLM training allows models to understand better by aligning model outputs more closely with human preferences and expectations. This approach is especially beneficial for CIP, where nuanced understanding and accurate, reliable communication are paramount. Adopting RLHF can enable more effective monitoring, threat detection, and incident response, thereby enhancing the resilience and security of critical infrastructure systems. Therefore, this adoption faces challenges such as defining appropriate reward functions that accurately reflect critical infrastructure systems' priorities and safety requirements.

\subsubsection{Quantization}
Quantization offers a pathway to reducing the computational demands of deploying LLMs in critical infrastructure settings. Several techniques exist for quantizing LLMs to 4-bit precision. Examples include QuaRot \cite{ashkboos2024quarot}, GPTQ \cite{frantar2022gptq}, AWQ \cite{lin2023awq}, SqueezeLLM \cite{kim2023squeezellm}, AQLM \cite{egiazarian2024extreme}, and llama.cpp with GGUF, all of which are well-regarded methods compatible with numerous frameworks. However, challenges arise in maintaining model accuracy and performance under reduced precision. Ensuring that the quantized models can reliably detect and respond to cyber threats without false positives or negatives is paramount. Balancing model size and computational efficiency with the need for real-time, high-performance decision-making in critical infrastructure contexts is a significant challenge.

\subsubsection{Retrieval-augmented generation (RAG)}
Integrating Retrieval Augmented Generation (RAG) technology into the CIP domain introduces a transformative approach to safeguarding vital assets such as power grids, water systems, and communication networks \cite{gao2023retrieval}. Unlike traditional language models that excel in general tasks but lack the depth for specialized applications, RAG's architecture—which combines an information retrieval component with a text generation model—perfectly aligns with critical infrastructure security's complex and dynamic nature. By enabling real-time access to external databases and the latest research on threats and vulnerabilities, RAG ensures the production of contextually relevant and factually accurate responses grounded in the most current information available. This capability is crucial for CIP, where rapidly assimilating and acting upon up-to-date intelligence can mean the difference between the regular operation of essential services and a potentially catastrophic failure.

The methodology introduced by Meta AI researchers \cite{lewis2020retrieval} can involve fine-tuning a pre-trained model with a comprehensive index of documents relevant to critical infrastructure, offers a tailored solution for enhancing threat intelligence, vulnerability assessments, and incident response strategies. For example, RAG can be leveraged to analyze threat actor tactics, techniques, and procedures (TTPs), assess the impact of potential vulnerabilities on critical systems, and generate informed recommendations for mitigating risks. The model’s strength in producing factual, specific, and diverse outputs significantly improves verifying facts and combating misinformation related to threats against critical infrastructure.

\subsubsection{Inference optimization}
The optimization of Generative AI and LLMs for inference in critical infrastructure is challenging. The techniques involve managing such models' extensive compute and memory requirements, including optimizing the attention mechanism and managing memory more effectively through batching \cite{yu2022orca, patel2023splitwise,li2023lightseq}, key-value caching \cite{kang2024gear}, and model parallelism \cite{shoeybi2019megatron,huang2019gpipe}. These optimizations are crucial for deploying LLMs in real-world applications, including critical infrastructure, where efficient, reliable, and fast processing of large volumes of data is essential. To apply these concepts to critical infrastructure, we need to focus on customizing model parallelism and memory management techniques to suit critical systems' specific needs and constraints, ensuring that LLMs can be used effectively without compromising the performance or security of these vital services.

\section{Future Directions}
\label{sec:future}
As CNIs evolve, adopting cutting-edge technologies becomes crucial to handle emerging threats and challenges. In this section, we explore future directions that promise to enhance the resilience and security of critical infrastructures.

One innovative approach to CNI protection is the use of digital twins. It offers a paradigm shift in critical infrastructure systems protection, management, and monitoring \cite{twinpot, twinPort}. Proactive mitigation measures can be implemented by simulating and evaluating the consequences of cyberattacks or system failures before their occurrence, thanks to the capability of digital twins. Virtual replicas of real assets and systems provide CNI stakeholders with previously unattainable insight into their operations, weaknesses, and possible sites of failure. Furthermore, by utilizing the sophisticated analytics and simulation capabilities of the digital twin-empower critical infrastructure, CNI operators can anticipate and reduce risks, improve performance, and simplify maintenance tasks. Integrating digital twin technology into CIP strategies promises to revolutionize the real-time monitoring, managing, and safeguarding of critical assets.

With the ability to perform complex calculations at speeds exponentially faster than classical computers, quantum computing promises to revolutionise threat detection, vulnerability assessments, and encryption methodologies. Quantum computing can enhance threat detection, risk analysis, and system optimisation for CIP. Critical infrastructure operators can maximise efficiency while minimising risks thanks to quantum algorithms' ability to solve optimisation issues like resource allocation and network routing. Additionally, massive databases can contain hidden patterns that quantum machine learning algorithms can find, providing proactive threat intelligence and flexible security solutions.  

With the rise of quantum computing, traditional encryption methods can be easily decrypted by quantum algorithms. Therefore, developing quantum-resistant encryption algorithms and cryptographic protocols is crucial to guarantee the long-term security of CNIs. Quantum encryption techniques provide secure communications and data transmission channels against even the most sophisticated cyber attacks \cite{2021Quantum}. Quantum encryption and cryptography methods have enormous potential to provide ultra-secure communication networks, protect critical infrastructures from malicious actors, and secure the confidentiality and integrity of sensitive information throughout CNIs.

Augmented reality (AR) technologies hold the potential to revolutionise operator interaction and visualisation of critical infrastructure systems. By placing digital data in the real world, AR improves situational awareness by enabling operators to identify irregularities and take immediate action in response to new threats. AR applications in CIP can give decision-makers immediate insights into operational status, security warnings, and system health, enabling them to make informed decisions in crisis moments. Moreover, AR-based training simulations can help staff gain practical experience, enhancing their preparedness and response skills.

Resilient and adaptive control systems are essential for maintaining operational functionality during disruptions as CNIs become increasingly interconnected and complex. 
Resilient control systems employ self-healing mechanisms, distributed control algorithms, and autonomous agents to effectively respond to changing conditions and mitigate the impact of cyberattacks and physical threats.  
Moreover, blockchain is another critical technology enhancing CIP. Blockchain's immutable and decentralized ledger presents special abilities to protect private information, confirm transactions, and guarantee the reliability of critical services\cite{2019Blockchain}. Stakeholders can develop a strong framework for securing sensitive infrastructure assets by utilizing blockchain technology for identity management, safe data sharing, and tamper-proof documentation of critical operations. Furthermore, smart contracts on blockchain systems can automate and enforce security measures, guaranteeing compliance with legal requirements and enhancing resilience against emerging threats \cite{2019IEEEIoTBlockchain}. 

The future of CIP is dependent on creativity, collaboration, and proactive security strategies. Through the integration of advanced technologies like digital twins, quantum encryption, and augmented reality, stakeholders can fortify critical infrastructure against constantly changing threats. By investing in their adoption as a strategic and progressive step, we can ensure that essential services continue for future generations.

\section{Conclusion}
\label{sec:conclusion}
As our society increasingly depends on networked digital systems, the possibility of cyberattacks targeting CNIs keeps also growing. CIP against cyberattacks is necessary to maintain the reliability and stability of critical services.
Addressing these challenges requires a diverse approach encompassing robust cybersecurity measures, comprehensive protection strategies, and continuous innovation. Throughout this study, we have explored the various challenges and solutions in protecting critical infrastructure, ranging from cybersecurity issues and privacy concerns to the potential deployment of advanced technologies like Generative AI and LLMs. CIP demands a proactive, multidisciplinary approach that integrates technological innovation, regulatory compliance, and collaborative partnerships. Protecting critical infrastructure requires large expenditures in strong cybersecurity defences, the strategic utilization of cutting-edge technology, and the development of a resilient organizational culture.





\bibliographystyle{IEEEtran}
\bibliography{ref} 

\begin{IEEEbiography}[{\includegraphics[width=1in,height=1.25in,clip,keepaspectratio]{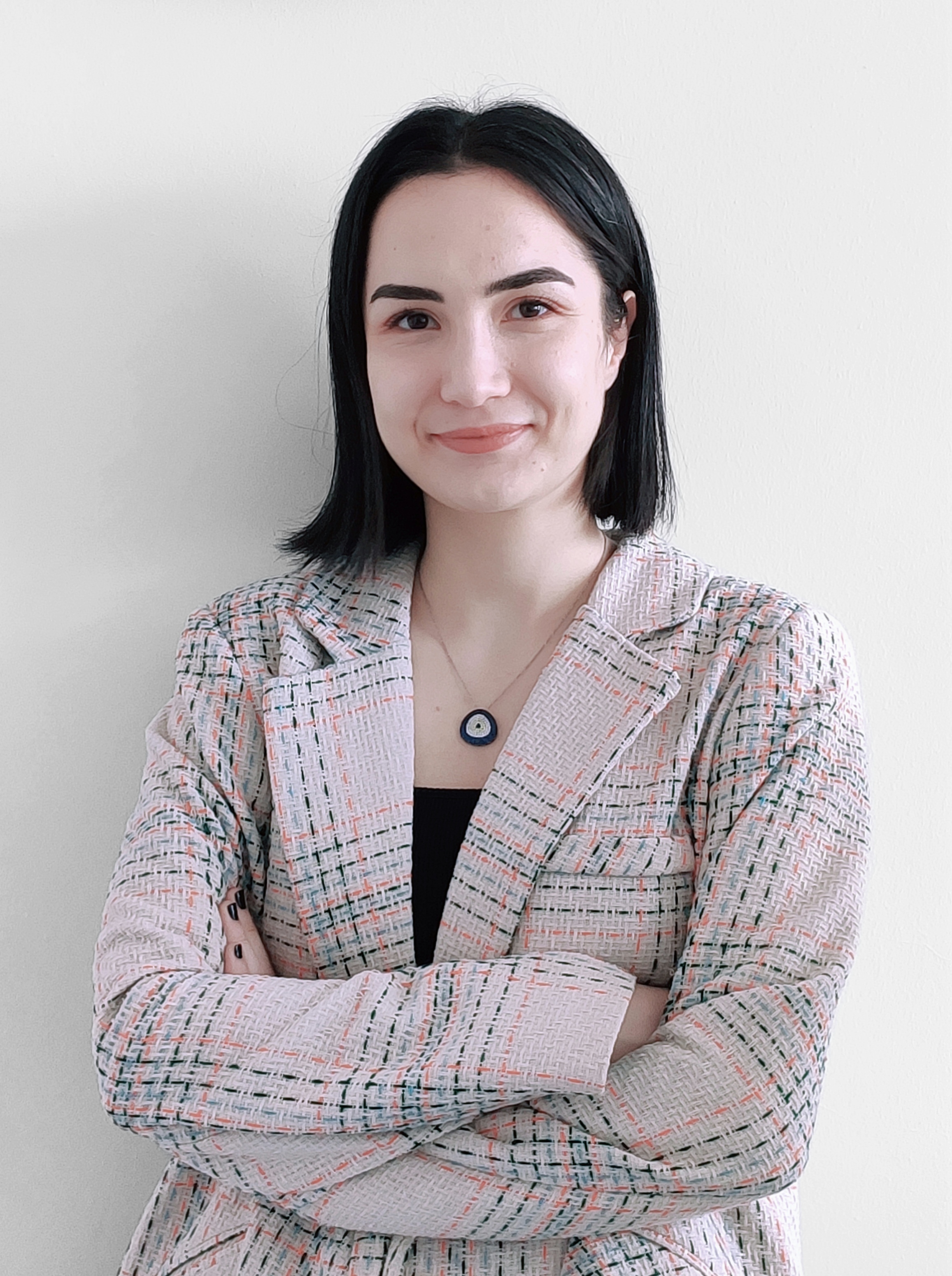}}]{YAGMUR YIGIT} is a PhD student in the School of Computing Engineering and The Built Environment. She received her MSc in the Department of Computer Engineering at the Istanbul Technical University and her BSc in Mechatronics Engineering from Istanbul Aydin University in 2023 and 2017, respectively. She worked as an R\&D engineer for two years at General Mobile, a Turkish smartphone company. Then, she worked as a 5G R\&D engineer for ten months at Netas, a Turkish company operating in information technologies. She was also one of the DeepMind scholars in 2020-2022 and a Turkcell Research Assistant in 2023. Her research focuses on AI-assisted next-generation cyber-secure networks.
\end{IEEEbiography}

\begin{IEEEbiography}[{\includegraphics[width=1in,height=1.25in,clip,keepaspectratio]{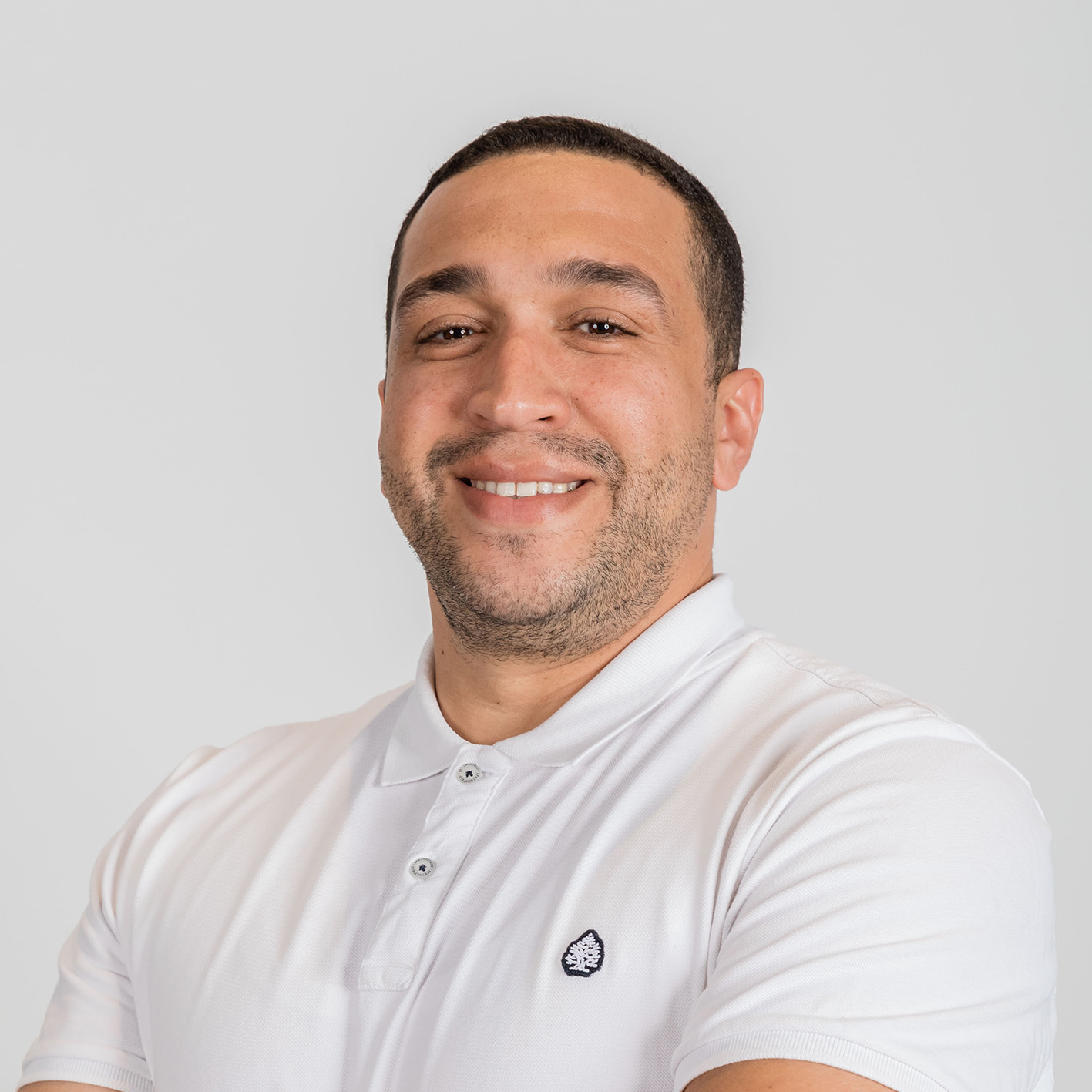}}]
 {MOHAMED AMINE FERRAG} earned his Bachelor's, Master's, Ph.D., and Habilitation degrees in Computer Science from Badji Mokhtar—Annaba University in Annaba, Algeria, completing his studies in 2008, 2010, 2014, and 2019, respectively. He served as an Associate Professor in the Department of Computer Science at Guelma University, Algeria, from 2014 until 2022. Concurrently, from 2019 to 2022, he held the position of Senior Researcher at the NAU-Lincoln Joint Research Center of Intelligent Engineering, based at Nanjing Agricultural University in China. As of 2022, Dr. Ferrag is the Lead Researcher at the Artificial Intelligence \& Digital Science Research Center at the Technology Innovation Institute in Abu Dhabi, United Arab Emirates. Dr. Ferrag's research is primarily focused on a spectrum of topics within the cyber security domain, including wireless network security, network coding security, applied cryptography, blockchain technology, generative AI, software security, and the application of AI in cyber security. His scholarly output includes over 140 papers published in international journals and conference proceedings. Dr. Ferrag has spearheaded numerous projects in research and development, fostering collaborative ties with academic institutions in the UK, Australia, USA, Canada, and China. His contributions to the field include the creation of three cybersecurity datasets, namely, Edge-IIoT dataset, CyberMetric dataset, and FormAI dataset, which have become essential resources for AI researchers worldwide. His academic contributions have been recognized with the 2021 IEEE TEM Best Paper Award, the 2022 Scopus Algeria Award, and many best paper conference awards. He has consistently been named on Stanford University's list of the world's top 2\% of scientists four times from 2020 through 2023. Dr. Ferrag also contributes to the academic community as an associate editor for prestigious journals, such as the IEEE Internet of Things Journal and ICT Express (Elsevier). In addition to his research and editorial roles, Dr. Ferrag is a Senior Member of the Institute of Electrical and Electronics Engineers (IEEE). 
\end{IEEEbiography}

\begin{IEEEbiography}[{\includegraphics[width=1in,height=1.25in,clip,keepaspectratio]{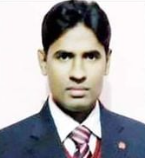}}]
 {Iqbal H. Sarker}  received his Ph.D. in Computer Science from Swinburne University of Technology, Melbourne, Australia in 2018. Now he is working as a
Research Fellow of the Cyber Security Cooperative Research Centre (CRC) in association with the Centre for Securing Digital Futures, Edith Cowan University
(ECU), Australia. His research interests include Cybersecurity, AI/XAI and Machine Learning Algorithms, Data Science and Behavioral Analytics, Trustworthy LLMs, Knowledge and Rule Mining, Digital Twin, Critical Infrastructures, Industry Applications. He has published 100+ journal and conference papers in various reputed venues published by Elsevier,
Springer Nature, IEEE, ACM, Oxford University Press, etc. Moreover, he is a lead author of the books "Context-Aware Machine Learning and Mobile Data
Analytics", and “AI-driven Cybersecurity and Threat Intelligence”, published by Springer Nature, Switzerland. He has also been listed in the world’s top 2\% of
most-cited scientists, published by Elsevier \& Stanford University, USA. In addition to research work and publications, Dr. Sarker is also involved in a number of research engagement and leadership roles such as Journal editorial, international conference program committee (PC), student supervision, visiting scholar
and national/international collaboration. He is a member of IEEE, ACM and Australian Information Security Association.
\end{IEEEbiography}

\begin{IEEEbiography}[{\includegraphics[width=1in,height=1.25in,clip,keepaspectratio]{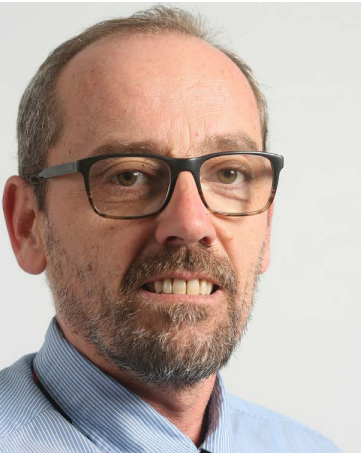}}]{Leandros A. Maglaras}  is a professor of cybersecurity in the School of Computing at Edinburgh Napier University. From September 2017 to November 2019, he was the Director of the National Cyber Security Authority of Greece. He obtained a B.Sc. (M.Sc. equivalent) in Electrical and Computer Engineering from the Aristotle University of Thessaloniki, Greece in 1998, MSc in Industrial Production and Management from the University of Thessaly in 2004, and M.Sc. and PhD degrees in Electrical \& Computer Engineering from the University of Thessaly, in 2008 and 2014 respectively. In 2018, he was awarded a PhD in Intrusion Detection in SCADA systems from the University of Huddersfield. He is featured in Stanford University's list of the world’s Top 2\% scientists. He is a Senior Member of the Institute of Electrical \& Electronics Engineers (IEEE) and is an author of more than 200 papers in scientific magazines and conferences.
\end{IEEEbiography}

\begin{IEEEbiography}
[{\includegraphics[width=1in,height=1.25in,clip,keepaspectratio]{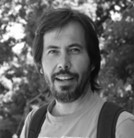}}]
{Dr. Christos Chrysoulas} received his Diploma and his Phd in Electrical Engineering from the University of Patras in 2003 and 2009 respectively. During his Phd (2004-2009) and PostDoc studies (2010-2015) his research was focused on Smart Grids, IoT, Industrial Automation, Machine Learning, Big Data, E-Learning systems, Computer Networks, High Performance Communication Subsystems Architecture and Implementation, Wireless Networks, Service Oriented Architectures (SOA), Resource Management and Dynamic Service Deployment in New Generation Networks and Communication Networks, Grid Architectures, Semantics, and Semantic Grid. He joined CISTER Research Center, Porto, as an invited Researched in 2013. He joined University of Porto as Post-Doc Research fellow in 2014 and from July 2015 he was with the University of Essex, holding a Senior Officer Researcher position. Currently he is holding an Assistant Professor’s position in Software Engineering and Big Data in Edinburgh Napier University, UK. 
The outcome of this effort was properly announced in more than 60 technical papers in these areas.  Dr. Christos Chrysoulas also participated as Senior Research/Engineer in both European and National Research Projects.
\end{IEEEbiography}

\begin{IEEEbiography}[{\includegraphics[width=1in,height=1.25in,clip,keepaspectratio]{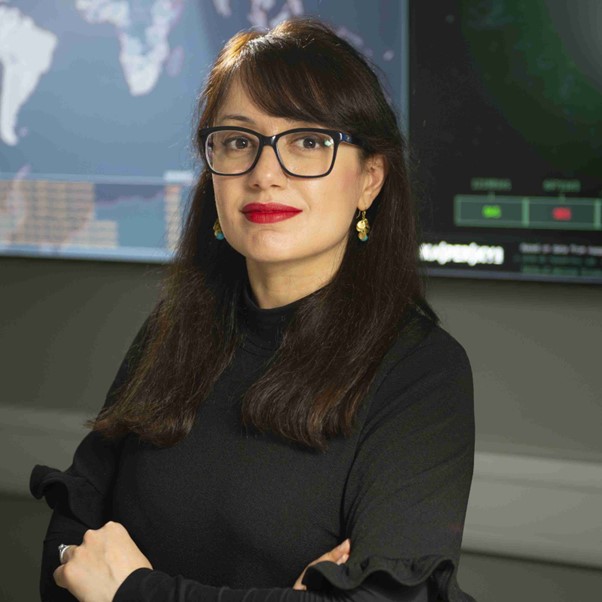}}]{Dr Naghmeh Moradpoor} (PhD, MSc, BSc, CEH, FHEA, SMIEEE) is a Lecturer in Cybersecurity and Networks, a core member of the Centre for Distributed Computing, Networking and Cybersecurity, and an active member of The Cyber Academy in the School of Computing, Engineering and the Built Environment (SCEBE) at Edinburgh Napier University. She is a Fellow of the Higher Education Academy (FHEA), a Senior Member of IEEE (SMIEEE), a Certified Ethical Hacker (CEH), and an active researcher since 2009. Naghmeh has a research background in machine learning for cybersecurity as well as next generation broadband access networks. This includes Industrial Control System \& Critical Infrastructure Protection security analysis \& countermeasures, detection and classification of various cybersecurity attacks for computer networks, data analytics \& decision support for cybersecurity, Intrusion Detection and Prevention Systems for Internet of Things, cyberbullying detection and prevention, and inter-vehicular trust in autonomous vehicles using Distributed Ledger Technology. Naghmeh was awarded Outstanding Woman in Cyber at the Scottish Cyber Awards 2017. These Awards are a highly prestigious and flagship event and the first of its kind in Scotland where the best of the best in cybersecurity are being recognised and awarded for their passion, dedication and hard work raising standards in both industry and academia. Naghmeh was also nominated for the Best Supervisor for Research Category by ENSA’s Student-Nominated Excellence Awards 2021. Naghmeh is in the top 2\% (top one million) most cited authors in the world. This is according to the World Scientist and University Rankings 2023 (AD Scientific Index 2023).Naghmeh is an Academic Editor for Journal of Security and Communication Networks, an Associate Editor for Journal of Ambient Intelligence and Humanized Computing (AIHC), and an Academic Editor for IET Information Security.
\end{IEEEbiography}

\begin{IEEEbiography}[{\includegraphics[width=1in, height=1.25in,clip, keepaspectratio]{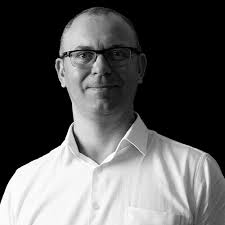}}]  {HELGE JANICKE} is currently the Research Director of the Cyber Security
Cooperative Research Centre, Australia. He is affiliated with Edith Cowan University and holds a visiting professorship in cyber security at De Montfort University, U.K. He established DMU’s Cyber Technology Institute and its Airbus Centre of Excellence in SCADA Cyber Security and Forensics Research. He has been the Head of the School of Computer Science, De Montfort University, U.K., before taking up his current position as the Research Director for the Cyber Security Cooperative Research Centre. He has founded the International Symposium on Industrial Control System Cyber Security Research (ICS-CSR) and contributed over 150 peer-reviewed articles and conference papers to the field that resulted from his collaborative research with industry partners such as Airbus, BT, Deloitte, Rolls-Royce, QinetiQ, and GeneralDynamics. His research interests include cyber security, in particular with applications in critical infrastructures using cyber-physical systems, SCADA, and industrial control systems. His current research investigates the application of agile techniques to cyber incident response in critical infrastructure, managing human errors that lead to cyber incidents, and research on cyber warfare and cyber peacekeeping.
\end{IEEEbiography}

\EOD

\end{document}